\documentclass[aps,twocolumn,showpacs,superscriptaddress,amsmath,amssymb,amsfonts,floatfix,longbibliography]{revtex4-2}

\usepackage[T1]{fontenc} 
\usepackage{graphicx}
\usepackage{dcolumn}
\usepackage{bm}
\usepackage{amssymb}
\usepackage{tabularx}
\usepackage{microtype}
\usepackage{xfrac}
\usepackage{chemformula}
\usepackage{gensymb}
\usepackage[most]{tcolorbox}
\usepackage{xcolor}
\usepackage{multirow}
\usepackage{enumitem}
\usepackage{array}
\usepackage{booktabs}
\usepackage[flushleft]{threeparttable}
\usepackage{natbib,hyperref}
\usepackage{xurl}
\usepackage{graphicx}
\usepackage{makecell}
\usepackage{xcolor}
\usepackage{soul}
\usepackage{lineno} 

\hypersetup{
	colorlinks=true, 
	citecolor=blue,  
}

\begin{document}

\title{Microscopic origin of magnetic order in spinel high entropy oxides}

\author{Mario Ulises Gonz\'alez-Rivas}
    \affiliation{Blusson Quantum Matter Institute, University of British Columbia, Vancouver, BC V6T 1Z4, Canada}
    \affiliation{Department of Physics \& Astronomy, University of British Columbia, Vancouver, BC V6T 1Z1, Canada}

\author{Chun-Fu Chang}
    \affiliation{Max Planck Institute for Chemical Physics of Solids, N\"othnitzer Stra\ss e 40, 01187, Dresden, Germany}

\author{Martin Bluschke}
      \affiliation{Blusson Quantum Matter Institute, University of British Columbia, Vancouver, BC V6T 1Z4, Canada}

\author{Jessica Freese} 
    \affiliation{Department of Physics and Engineering Physics, University of Saskatchewan, Saskatoon, SK S7N 5E2, Canada}

\author{Peter Bencok} 
    \affiliation{Diamond Light Source, Harwell Science and Innovation Campus, Didcot, OX11 0DE, United Kingdom}

\author{Ronny Sutarto} 
    \affiliation{Canadian Light Source, Saskatoon, Saskatchewan S7N 2V3, Canada}

\author{Teak D. Boyko} 
    \affiliation{Canadian Light Source, Saskatoon, Saskatchewan S7N 2V3, Canada}

\author{Robert J. Green}
    \affiliation{Department of Physics and Engineering Physics, University of Saskatchewan, Saskatoon, SK S7N 5E2, Canada}

    \affiliation{Blusson Quantum Matter Institute, University of British Columbia, Vancouver, BC V6T 1Z4, Canada}

\author{George A. Sawatzky}
    \affiliation{Blusson Quantum Matter Institute, University of British Columbia, Vancouver, BC V6T 1Z4, Canada}
    \affiliation{Department of Physics \& Astronomy, University of British Columbia, Vancouver, BC V6T 1Z1, Canada}

\author{Liu Hao Tjeng}
    \affiliation{Max Planck Institute for Chemical Physics of Solids, N\"othnitzer Stra\ss e 40, 01187, Dresden, Germany}

\author{Alannah M. Hallas}
\email[Email: ]{alannah.hallas@ubc.ca}
    \affiliation{Blusson Quantum Matter Institute, University of British Columbia, Vancouver, BC V6T 1Z4, Canada}
    \affiliation{Department of Physics \& Astronomy, University of British Columbia, Vancouver, BC V6T 1Z1, Canada}
    \affiliation{Canadian Institute for Advanced Research (CIFAR), Toronto, ON, M5G 1M1, Canada}

\date{\today}
\begin{abstract}

High entropy oxides (HEOs) can possess long-range ordered magnetic states despite their extreme chemical disorder. Very little is known about how the different chemical constituents in HEOs contribute to the emergence of these magnetic states. In this work, we leverage element-specific magnetometry attained via x-ray magnetic circular dichroism (XMCD) to understand how magnetic order is driven in two ferrimagnetic spinel-structured HEOs with compositions \ch{(Cr,Mn,Fe,Co,Ni)3O4} and \ch{(Cr,Mn,Fe,Co,Ni)_{2.4}Ga_{0.6}O4}. 
We find that while the magnetic transition is simultaneous for all chemical species, the rate at which their magnetic moments grow is strongly cation dependent. This behavior is explained by the varying $\textit{3d}$ crystal field level fillings of the magnetic cations, which in turn determine their ability to participate in the different magnetic exchange pathways available in the spinel structure. Dominant $A$-$B$ sublattice exchange enables some species to harden rapidly while others exhibit a sluggish transition due to frustration from competing interactions. 
Non-magnetic substitution suppresses these differences, introducing broken magnetic linkages that relieve frustration. These observations provide a blueprint for the rational design of magnetic HEOs to translate their chemical complexity into targeted functionality.

\end{abstract}

\maketitle

\section*{\label{sec:Introduction}Introduction}

First established in 2015~\cite{rost2015entropy}, high entropy oxides (HEOs) are metal oxide systems where equiatomic ratios of at least five cations are randomly distributed across one of the material's sublattices~\cite{brahlek2022name}. On average, the resulting oxide preserves the long-range order and corresponding space group symmetry of a conventional crystalline material, whilst locally, the identity of a cation will be random. As a result, these systems are characterized by a large configurational contribution to their entropy. 
Despite their profound disorder, HEOs exhibit complex collective phenomena, including ferroelectricity~\cite{sharma2022high,qi2023high}, metal-to-insulator transitions~\cite{cui2023orbital}, and magnetism~\cite{sarkar2021magnetic}.

Magnetism provides an ideal framework for probing collective phenomena in HEOs. Just as cation disorder creates a highly random chemical environment, magnetic HEOs exhibit similarly non-uniform and potentially competing magnetic exchange interactions. Despite this complexity, long-range magnetic order has been observed in a variety of HEOs, including those with rocksalt~\cite{Zhang2019rocksaltAFM,jimenez2019long}, perovskite~\cite{witte2019high,witte2020magnetic,Mazza2022HEOPerovskitedesign}, and spinel~\cite{Musico2019TunableSpinel,johnstone2022entropy,nevgi2025local} structures. These materials have been found to adopt relatively simple antiferromagnetic or ferrimagnetic states, with long correlation lengths evidenced by resolution-limited magnetic Bragg peaks~\cite{Zhang2019rocksaltAFM,johnstone2022entropy}. 
How these simple magnetic orders emerge from underlying disorder remains unclear, as does the extent to which the ordered states are glassy or dynamic.

\begin{figure*}[htbp]
\includegraphics[width=1.3\columnwidth]{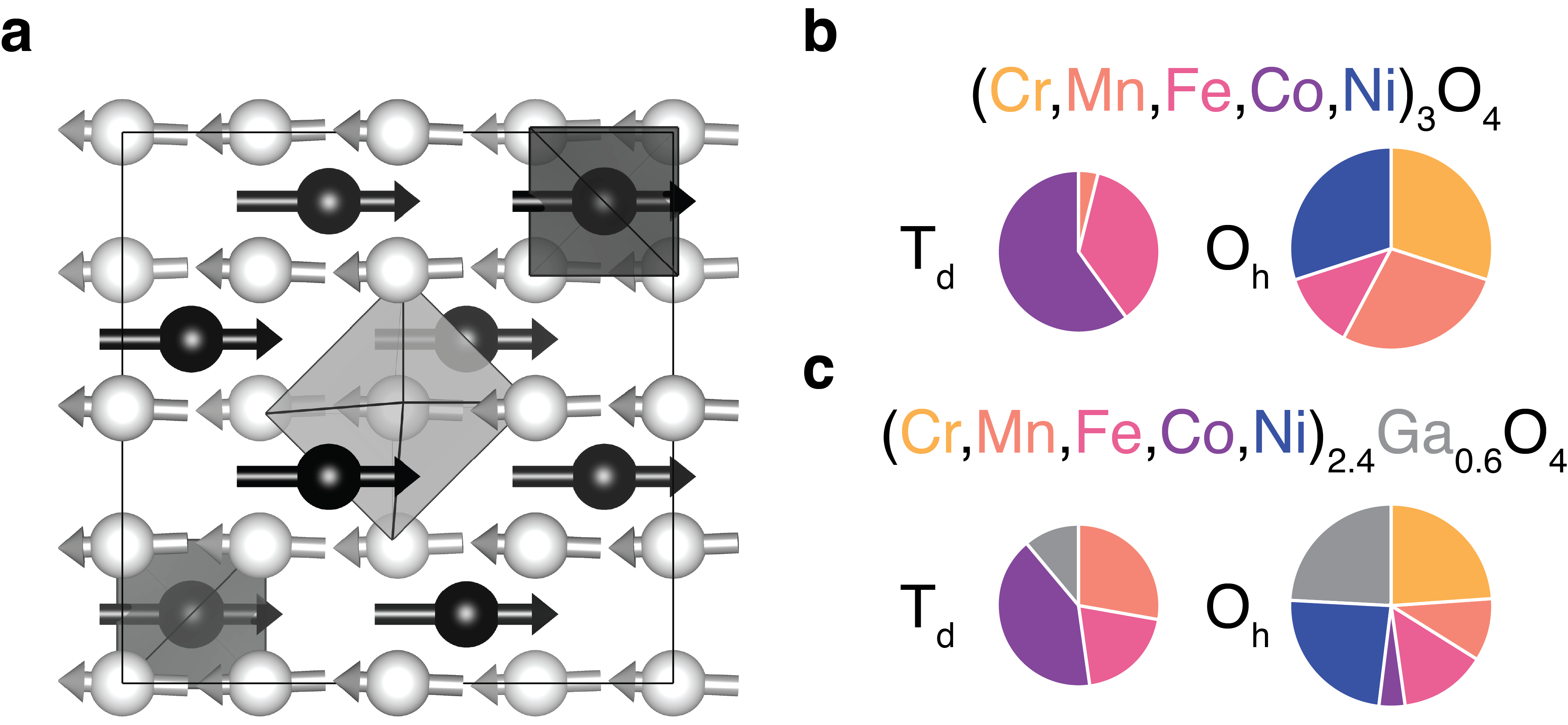}
\caption{\textbf{Magnetic structure and cation distribution of ferrimagnetic spinel HEOs.} \textbf{a} \ch{(Cr,Mn,Fe,Co,Ni)3O4} and \ch{(Cr,Mn,Fe,Co,Ni)_{2.4}Ga_{0.6}O4} are cubic spinels with ferrimagnetic order. The spinel structure contains two distinct cation sites, one with octahedral ($O_h$) oxygen coordination and the other tetrahedral ($T_d$), shown in white and black, respectively. The magnetic moments on the two sublattices are coupled antiferromagnetically but are unequal in magnitude, yielding a ferrimagnetic ordered state with a small canting of the octahedral moments. Site selectivity-driven cation distributions in \textbf{b} \ch{(Cr,Mn,Fe,Co,Ni)3O4} and \textbf{c} \ch{(Cr,Mn,Fe,Co,Ni)_{2.4}Ga_{0.6}O4}, where the areas of the pie charts are scaled to represent the 2:1 ratio of $O_h$ to $T_d$ sites.} 
 \label{fig:magstructure}
\end{figure*} 

Developing a complete picture of collective behavior in HEOs requires direct knowledge of the roles played by the constituent cations.
At the structural level, great insights 
have been achieved using spectroscopic probes with elemental sensitivity, such as 
x-ray absorption spectroscopy (XAS), 
and mass-sensitive techniques, such as atom probe tomography. These have led to important insights, such as establishing atomic randomness in the prototypical rocksalt HEO~\cite{chellali2019homogeneity}, uncovering strong site selectivity in spinel HEOs~\cite{sarkar2022comprehensive,johnstone2022entropy,masina2025multi,Meier2025-HEO-XAS-RIXS}, and the detection of element-specific local 
distortions~\cite{nevgi2025local}.
These studies have also yielded important insights into HEO functionality~\cite{wang2023synergy}. 
Beyond crystal structure, disentangling the contributions of individual cations to other collective properties of HEOs remains a formidable challenge. Most experimental probes of magnetism lack direct elemental sensitivity, so measured properties generally reflect an average over all constituent elements.

In this paper, we exploit x-ray magnetic circular dichroism's (XMCD) element specific magnetometry capabilities to disentangle cation-specific, temperature-dependent magnetic moments in two ferrimagnetic HEOs with the spinel structure. 
Through temperature-dependent XMCD, we study the elementally resolved onset and evolution of magnetic order in these HEOs. 
The magnetic transition occurs simultaneously for all cations in \ch{(Cr,Mn,Fe,Co,Ni)3O4}, their individual magnetic moments exhibit a rich phenomenology at lower temperatures. This richness can be explained in terms of the different magnetic exchange pathways present in the spinel structure, whose availability varies as a function of crystal field symmetry and \textit{3d} orbital filling, which results in specific cations experiencing magnetic frustration as they try to satisfy competing interactions, while others only engage in synergistic exchange channels. Partial substitution of non-magnetic Ga in \ch{(Cr,Mn,Fe,Co,Ni)_{2.4}Ga_{0.6}O4} weakens these differences, as the broken magnetic linkages provide relief from magnetic frustration.

\section*{Results}

\begin{table*}[htbp]
    \centering
\caption{ \textcolor{black}{Reported cation distributions in \ch{(Cr,Mn,Fe,Co,Ni)3O4} from the literature, along with the probes employed, which in addition to XAS and XMCD, include M\"ossbauer, x-ray absorption near edge structure (XANES), resonant inelastic x-ray scattering (RIXS), and resonant x-ray diffraction (ReXRD). In the chemical formula, ( ) denotes the spinel structure's tetrahedral site, and [ ] the octahedral site. }}

    \begin{tabular}{lcc}
        \toprule
         
          & \textcolor{black}{Probes} & \textcolor{black}{Chemical formula}\\
         \hline
         \textcolor{black}{Ref.~\cite{sarkar2022comprehensive}} & \textcolor{black}{XAS, XMCD, XANES, M\"ossbauer} & \textcolor{black}{\ch{(Fe_{0.4}Co_{0.6})[Cr_{0.6}Mn_{0.6}Fe_{0.2}Ni_{0.6}] O4}} \\
         \textcolor{black}{Ref.~\cite{johnstone2022entropy}} & \textcolor{black}{XAS, XMCD} &  \textcolor{black}{\ch{(Mn_{0.06}Fe_{0.33}Co_{0.61}) [Cr_{0.61} Mn_{0.55} Fe_{0.28}Ni_{0.54}]O4}} \\
         \textcolor{black}{Ref.~\cite{Meier2025-HEO-XAS-RIXS}} & \textcolor{black}{XAS, RIXS} &  \textcolor{black}{\ch{(Mn_{0.12}Fe_{0.46}Co_{0.39})[Cr_{0.6}Mn_{0.48}Fe_{0.14}Co_{0.21}Ni_{0.6}]O4}}\\
         \textcolor{black}{Ref.~\cite{masina2025multi}} & \textcolor{black}{XANES, ReXRD} & \textcolor{black}{\ch{(Mn_{0.10}Fe_{0.41}Co_{0.49}) [Cr_{0.66}Mn_{0.58}Fe_{0.22}Co_{0.04}Ni_{0.5}]O4}$^*$}\\
         \textcolor{black}{This work} & \textcolor{black}{XAS, XMCD} & \textcolor{black}{\ch{(Mn_{0.04}Fe_{0.36}Co_{0.6}) [Cr_{0.6}Mn_{0.56}Fe_{0.24}Ni_{0.6}]O4}} \\
         \toprule
    \end{tabular}\\
    \raggedright \footnotesize \textcolor{black}{$^{*}$This composition does not conform to the expected equimolarity of the nominal composition. \\}
    \label{table:composition}
\end{table*}

\subsection{Crystal and magnetic structures of spinel HEOs}

Phase pure, polycrystalline samples of spinel HEOs with nominal composition \ch{(Cr,Mn,Fe,Co,Ni)_{3-x}Ga_{x}O4} (x = 0, 0.6) were synthesized via combustion synthesis, \textcolor{black}{followed by an annealing cycle,} adapting the methodology described in Ref.~\cite{JACS-synthesis-dep}. Structural characterization is shown in the SI. These materials crystallize in a cubic spinel structure (space group $Fd\overline{3}m$) as shown in Fig.~\ref{fig:magstructure}(a). 
Spinels, oxides with the nominal chemical formula $AB_2$O$_4$, are characterized by the presence of two cation sublattices with different anion coordination –– the \textit{A} site is tetrahedral ($T_d$), and the \textit{B} site octahedral ($O_h$). The two most important magnetic exchange pathways are the edge-sharing links between octahedral sites and the corner-sharing links between the two sublattices.  
These high entropy spinels exhibit a slightly canted ferrimagnetic order (Fig.~\ref{fig:magstructure}(a)) where the octahedral and tetrahedral sites are antiferromagnetically coupled and uncompensated, resulting in a non-zero net magnetic moment~\cite{Spinel-HEO-Dabrowa,Spinel-HEO-Cieslak-FiM-Mossbauer,Spinel-HEO-Mao-FiM,sarkar2022comprehensive}. They exhibit high Curie temperatures ($T_{\textrm{C}}$) with the fully magnetic \ch{(Cr,Mn,Fe,Co,Ni)3O4} ordering at $T_{\textrm{C}}\approx400$~K, and the Ga-substituted \ch{(Cr,Mn,Fe,Co,Ni)_{2.4}Ga_{0.6}O4} having $T_{\textrm{C}}\approx300$~K~\cite{johnstone2022entropy}. 

The octahedral and tetrahedral sites in the spinel have inverted crystal field energy level splittings. As a result, there is a strong energetic preference for some cations (based on their $\textit{3d}$ electron fillings) to preferentially occupy one of the two sublattices. These effects lead to high entropy spinels exhibiting site selectivity, where the cations are not distributed randomly across both sublattices~\cite{sarkar2022comprehensive,johnstone2022entropy,masina2025multi,Meier2025-HEO-XAS-RIXS}. 
Fig.~\ref{fig:magstructure}(b,c) displays the site occupancies for \ch{(Cr,Mn,Fe,Co,Ni)3O4} and \ch{(Cr,Mn,Fe,Co,Ni)_{2.4}Ga_{0.6}O4}, as determined from XAS and XMCD, as described in the following section. \textcolor{black}{The actual cation distributions in these high entropy spinels are 
\ch{(Mn_{0.04}Fe_{0.36}Co_{0.6}) [Cr_{0.6}Mn_{0.56}Fe_{0.24}Ni_{0.6}]O4} and 
 \ch{(Mn_{0.28}Fe_{0.20}Co_{0.41}Ga_{0.11})[Cr_{0.48}Mn_{0.20}Fe_{0.28}Co_{0.07}Ni_{0.48}\allowbreak Ga_{0.49}]O4} where ( ) denotes cations occupying the spinel structure's tetrahedral sites, and [ ] those occupying the octahedral sites. The site assignments for the widely studied composition \ch{(Cr,Mn,Fe,Co,Ni)3O4} are in good agreement with the most recent literature on these materials, as shown in Table~\ref{table:composition}. }

\textcolor{black}{A previous study showed minimal dependence of the local and long-range structures in this family of spinel HEOs as a function of the synthesis route~\cite{JACS-synthesis-dep}. In terms of the cation distribution, the only changes observed were minimal variations in the Mn oxidation state, effectively corresponding to different levels of oxygen deficiency and a minimal charge doping of the structure. 
Likewise, there were minimal differences in the magnetic response of materials obtained via different methods. Therefore, and as magnetic interactions in localized systems are dominated by local structure and short-range interactions, our findings should be generalizable to spinel HEOs regardless of the synthesis routes.}

\subsection{Determination of cation occupancies and valences from XAS}\label{XAS_disc}

\begin{figure*}[htbp]
\includegraphics[width=\textwidth]{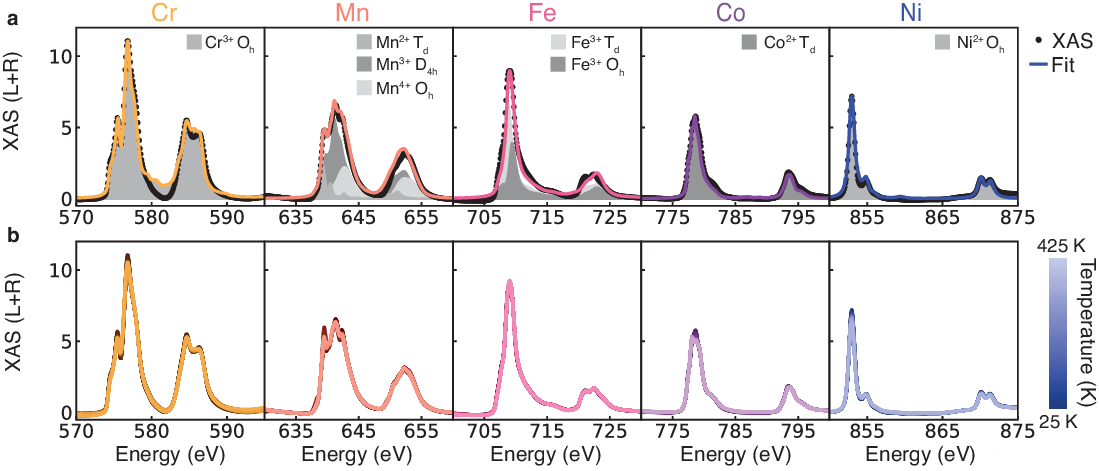}
\caption{\textbf{Temperature-dependent x-ray absorption spectroscopy (XAS)  
of \ch{(Cr,Mn,Fe,Co,Ni)3O4}.} \textbf{a} Ligand field multiplet theory fits to the 25~K XAS (L+R) at the $L_{2,3}$ edges for each of the cations present in the sample. Experimental data are presented as black dots and the optimized fits as colored lines. The areas of individual components are presented in shades of gray for each cation species. \textbf{b} Temperature dependence of the XAS (L+R) for each cation between 25~K and 425~K. Each spectrum is presented as a solid line, with its temperature encoded in the luminance of its color, following the scale on the right. Minimal temperature dependence is observed in the XAS, reflecting the stable cation valences. }
 \label{fig:XAS}
\end{figure*}

Our trek towards disentangling the roles that individual cations play in the emergence of magnetic order in HEOs begins with x-ray absorption spectroscopy. 
As the prototypical magnetic spinel HEO, several efforts have been made to quantify the cation distribution in \ch{(Cr,Mn,Fe,Co,Ni)3O4}~\cite{sarkar2022comprehensive,johnstone2022entropy,he2023selective,Meier2025-HEO-XAS-RIXS,masina2025multi}. While there exists a consensus for cations with the strongest site preferences –– Cr and Ni occupy octahedral sites exclusively –– disagreement remains regarding how Mn, Fe, and Co distribute across the remaining available sites~\cite{sarkar2022comprehensive,Meier2025-HEO-XAS-RIXS,johnstone2022entropy}. 

In this work, we build upon previous efforts by employing simultaneous analysis of the XAS and XMCD signals. They were modeled using ligand field multiplet theory (LFMT) calculations. The calculations were performed using the XTLS 9.0 code~\cite{Tanaka1994XTLS}. 
As dichroism is not only sensitive to any imbalance in spins, but also their direction relative to the incident polarization, additional contrast is gained between the octahedral and tetrahedral sites, which exhibit XMCD signals with opposite signs.
Since the crystal fields are then set to be the same for both the XAS and XMCD simulations, the fitted parameters are better constrained, and better resolution for chemical and electronic speciation is attained. Our discussion initially focuses on the fully magnetic \ch{(Cr,Mn,Fe,Co,Ni)3O4} parent composition. Once general trends and observations are established, we extend our findings to the magnetically diluted case.

We begin by establishing the site occupancies and valences of the constituent elements in \ch{(Cr,Mn,Fe,Co,Ni)3O4} in the low temperature limit. Total electron yield (TEY) XAS (L+R, where L and R stand for left circular and right circular polarization, respectively ) data, collected at 25~K under an applied field of 1.5 T, for the  $L_{2,3}$ edges of Cr, Mn, Fe, Co, and Ni in that order, are presented in Fig.~\ref{fig:XAS}(a) represented by black dots, with the simulated spectra presented as colored continuous lines, and their individual components as shaded areas. The corresponding XMCD (L$-$R) data and fits are shown in Fig.~\ref{fig:XMCD} (a). As previously established, the XAS and XMCD of \ch{Cr^3+} and \ch{Ni^{2+}} are unequivocally characteristic of purely $O_h$ coordination, which follows naturally from their large octahedral site preference energies.
Two \ch{Fe^{3+}} species are present, occupying the $O_h$ and $T_d$ sites in a 40:60 distribution, consistent with minimal site selectivity for a $d^5$ electron configuration. Co appears as \ch{Co^{2+}} in $T_d$ coordination, although a small ($\leq5\%$) \ch{Co^{3+}} $O_h$ contribution cannot be ruled out. Finally, Mn presents the richest crystal chemistry, appearing in three different species (as defined by their joint oxidation state and coordination environment): \ch{Mn^{2+}} in $T_d$ coordination, \ch{Mn^{3+}} in a tetragonally-distorted octahedral environment with $D_{4h}$ point-group symmetry, and \ch{Mn^{4+}} $O_h$. However, the $T_d$ fraction is minimal, on the order of $\sim5\%$. This assignment is consistent with the overall 2:1 site distribution of the spinel structure and also respects overall charge neutrality with negligible oxygen non-stoichiometries.

Temperature-dependent XAS spectra at the transition metal $L_{2,3}$ edges for \ch{(Cr,Mn,Fe,Co,Ni)3O4} between 25 K and 425 K are presented in Fig.~\ref{fig:XAS}(b). We can immediately observe that for Cr, Fe, and Ni, the XAS presents minimal temperature dependence. Indeed, for most transition metals, one would not expect any excited states to become thermally populated within our measurement range –– the typical energy scale for $d-d$ excitations is on the order of $10^{3}-10^4$~K. 

For Co and Mn, more significant spectral weight transfer is present as a function of temperature (Fig.~S3). In Co, weight is transferred from a feature located at 778.6 eV to one at 778.0 eV. To rule out surface chemical activity, we performed bulk-sensitive inverse partial fluorescence yield (IPFY) XAS at 20 K and 300 K (Fig. S3 (\textbf{d})), which shows the same spectral weight redistribution as the TEY measurement. We therefore assign this feature as originating from the thermal population of low-lying multiplets, facilitated by \ch{Co^{2+}} $T_d$'s weak field, ground state orbital degeneracy, and strong spin-orbit coupling~\cite{TanabeSugano1954,atkins2010shriver}. Additional evidence supporting this interpretation can be observed in the Co L-edge resonant inelastic x-ray scattering~\cite{Meier2025-HEO-XAS-RIXS}, where broadening of the elastic line and several low-lying peaks are observed, both suggesting the presence of an excited crystal field state. For Mn, the high-temperature XAS is characterized by an enhancement of a feature at 639.5 eV, which is mainly associated with \ch{Mn^{2+}} in a $T_d$ site, whilst the ratio of \ch{Mn^{3+}} and \ch{Mn^{4+}} –– in octahedral-like $D_{4h}$ and $O_h$ environments, respectively –– remains largely unchanged. This time, comparison with IPFY data collected at 20 K and 300 K (Fig. S3 (\textbf{a,b})) reveals no difference for Mn $L_{2,3}$, indicating that the variations observed in TEY originate from surface chemistry ––  likely enabled by Mn's low redox potential –– as opposed to a bulk effect.

\subsection{Cation-dependent magnetic moments obtained from XMCD analysis}\label{sumrules_disc}

\begin{figure*}[htbp]
\includegraphics[width=\textwidth]{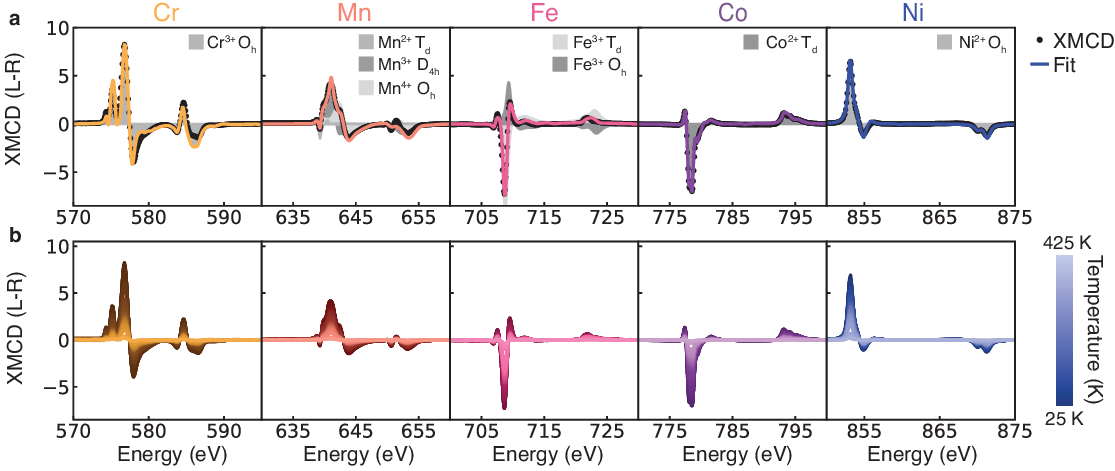}
\caption{\textbf{Temperature-dependent x-ray  magnetic circular dichroism (XMCD) of \ch{(Cr,Mn,Fe,Co,Ni)3O4}.} \textbf{a} Ligand field multiplet theory fits to the 25~K XMCD (L$-$R) at the $L_{2,3}$ edges for each of the cations present in the sample. Experimental data are presented as black dots and the optimized fits as colored lines. The areas of individual components are presented in shades of gray for each cation species.  \textbf{b} Temperature dependence of the  XMCD (L$-$R) for each cation between 25~K and 425~K. Each spectrum is presented as a solid line, with its temperature encoded in the luminance of its color, following the scale on the right. There is a continuous suppression in the XMCD intensity due to the decreasing ordered magnetic moment approaching the Curie temperature. }
 \label{fig:XMCD}
\end{figure*}

We are now in a good position to interpret the $L_{2,3}$ XMCD of \ch{(Cr,Mn,Fe,Co,Ni)3O4} and its temperature dependence, starting at a qualitative level. The complete XMCD dataset is shown in Fig.~\ref{fig:XMCD}(b). As expected, the XMCD intensity for all cations decreases with increasing temperature, a result consistent with their spin moments demagnetizing as thermal fluctuations begin to dominate. Furthermore, for several cations, a loss of fine structure in the XMCD signal becomes apparent at the highest temperatures. These changes indicate a breakdown of magnetic order as the ions approach the Curie temperature ($T_{\textrm{C}} \approx 400 K$).

\begin{figure*}[tb]
\includegraphics[width=\textwidth]{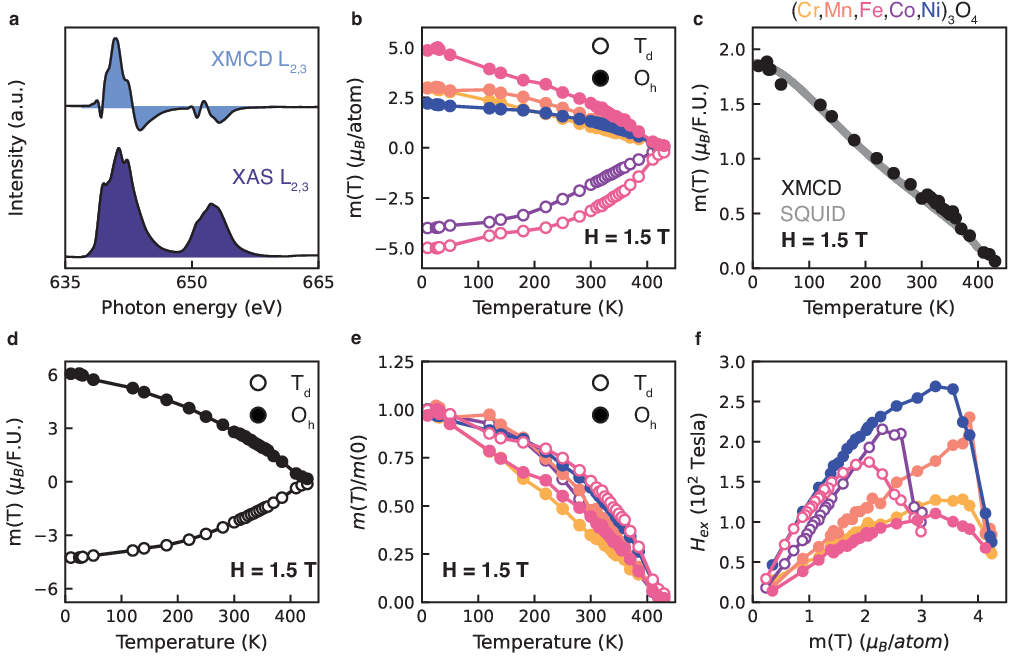}
\caption{\textbf{Magnetic moments extracted from XMCD in \ch{(Cr,Mn,Fe,Co,Ni)3O4}.} \textbf{a} Example XAS and XMCD $L_{2,3}$ spectra, highlighting their respective areas, of which XMCD/XAS is proportional to a cation's magnetic moment. \textbf{b} Temperature-dependent, element-specific magnetic moments for \ch{(Cr,Mn,Fe,Co,Ni)3O4}. A variety of hardening rates and saturated moments are observed. \textbf{c} Comparison of the total magnetic moment in \ch{(Cr,Mn,Fe,Co,Ni)3O4} as determined via XMCD with the bulk response from SQUID magnetometry under equivalent conditions ($H = 1.5$~T). Excellent agreement is observed across the temperature range. \textbf{d} Temperature-dependent magnetic moments from XMCD, decomposed into their octahedral and tetrahedral components. The $O_h$ sublattice's moment remains larger at all temperatures. \textbf{e} Element-specific magnetic moments in \ch{(Cr,Mn,Fe,Co,Ni)3O4}, as a function of temperature, normalized relative to their base temperature values, clearly demonstrating the variation in ordering rates for the different elements. \textbf{f} Exchange field felt by each cation, as a function of the opposite sublattice's magnetic moment. For smaller sublattice moments (higher temperatures), a nearly linear relationship is observed, a behavior that breaks down at larger sublattice magnetizations (lower temperatures), as the magnetization saturates and fluctuations become more prominent.}
 \label{fig:momenta}
\end{figure*}

Our multiplet theory models allow for the direct calculation of ground state properties of each crystallochemical species present in \ch{(Cr,Mn,Fe,Co,Ni)3O4}, including their spin and orbital angular momenta. $\langle S_z\rangle$ and $\langle L_z \rangle$ were computed, then used to compute the resulting magnetic moment, corresponding to each cation at full polarization. Once these values were extracted, the relative XMCD/XAS magnitude of the experimental data were compared to that of the calculations, to determine the fraction of the moment that the experimental XMCD attains at a given temperature, and extract element and site-specific magnetic moments ~\cite{VanderLaan2014XMCDreview}. The XMCD/XAS ratio is represented graphically in Fig.~\ref{fig:momenta}(a)

We first examine the temperature dependence of the absolute values of the magnetic moments for each of the cations in \ch{(Cr,Mn,Fe,Co,Ni)3O4}, which are presented in Fig.~\ref{fig:momenta}(b). The first level of categorization is provided by the sign of the magnetic moment for each cation: $O_h$ Cr, Mn, \ch{Fe^{3+}}, and Ni have positive net moments, whilst $T_d$ \ch{Fe^{3+}} and Co are net negative. Accordingly, the two cation subsets, identified from the dichroism's sign, correspond perfectly to those located primarily in octahedral and tetrahedral environments. Furthermore, every cation's moment goes towards $0\;\mu_B$ as the temperature approaches $T_{\textrm{C}}$, as expected for cations demagnetizing as they approach the bulk magnetic transition. Within experimental resolution, $T_C$ is observed to be the same for all cations.

The magnitudes of the low-temperature moments can be compared to the expected moment for each transition metal's electron configuration, as calculated from the electron count deduced from the XAS speciation. The results of this analysis are presented in Table~\ref{table:critexponents}. Agreement with the spin-only moments is excellent for the earlier transition metals (Cr, Mn, and Fe), but Co and Ni present significant orbital angular momentum contributions. In the case of Co, the large orbital contribution to its angular momentum ($\langle L_z \rangle = 0.47$), results in a higher magnetic moment than would be expected from a spin-only calculation (4.0 vs 3.0 $\mu_B$). The same is true of Ni, where $\langle L_z\rangle = 0.32$, resulting in a total moment of 2.2 $\mu_B$. 

We next validate our XMCD magnetic moments moments 
by comparing the data to the bulk magnetic moment under equivalent conditions, at $H = 1.5$~T. In Fig.~\ref{fig:momenta}(c) we compare the DC magnetic moment of \ch{(Cr,Mn,Fe,Co,Ni)3O4}, measured using SQUID magnetometry, to that recovered by adding the individual XMCD magnetic moments at each temperature step. Overall, the two signals show remarkable qualitative and quantitative agreement, with the total XMCD moment remaining within 10\% of the SQUID data for all temperatures below 330~K, with no adjustable parameters.
It is worth noting that the ferrimagnetic transition observed in low field susceptibility (see Fig.~S2) becomes appreciably broadened at the higher fields required for XMCD, which may further accentuate differences between the cations. The total tetrahedral and octahedral moments per formula unit are plotted in Fig.~\ref{fig:momenta}(d), where it can be observed that the octahedral moment is larger throughout the entire ordered regime.

As the XMCD magnetic moments extracted for \ch{(Cr,Mn,Fe,Co,Ni)3O4} are physically and chemically reasonable, we now consider their temperature dependence in more detail. The 25 K values represent the maximum observed moment for all cations, which can be inferred as the bulk magnetic moment flattens below 50 K. Therefore, it is most informative to look at the XMCD moments normalized to their maximum values, as presented in Fig.~\ref{fig:momenta}(e), which is effectively an elementally resolved order parameter. Significant qualitative differences are immediately apparent.
In the vicinity of the magnetic transition, Fe $T_d$ and Ni order the fastest of all the species present, with the temperature dependence of the two species' moments closely tracking one another.
In the opposing limit, Cr stands out, as its magnetic ordering lags behind the other constituents across the full measured temperature range. 
It is characterized by a nearly linear growth throughout the measured range. Similar behavior arises for Fe occupying $O_h$ sites, which, below 300~K, correlates strongly with Cr.

To parametrize the rate of magnetic hardening of the various cations -- away from $T_C$ -- we define the hardening temperature $T_H$ as the temperature at which the moment reaches 85\% of its final saturation value. The tabulated $T_H$ values for \ch{(Cr,Mn,Fe,Co,Ni)3O4}, shown in Table~\ref{table:critexponents}, effectively separate the cations into two categories: \ch{Cr^{3+}} and \ch{Fe^{3+}} $O_h$, which exhibit the aforementioned sluggish, nearly linear ordering rate, for which $T_H \leq 100$ K, and \ch{Mn^{3+/4+}}, \ch{Fe^{3+}} $T_d$, and \ch{Co^{2+}}, species with $T_H \geq 150$ K, with a faster rate of growth for their magnetic moments.

\begin{table}
    \centering
\caption{Cation-resolved XMCD magnetic moments ($m$) at 25 K, hardening temperatures ($T_H$), and exchange constants $J$, in \ch{(Cr,Mn,Fe,Co,Ni)3O4}. \textcolor{black}{Errors on $T_H$, $J_{AB}$, and $J_{BB}$ are fitting uncertainties. Errors on the observed XMCD total moment are the propagation of uncertainty from the site assignments of Mn and Fe cations, which carry greater uncertainty due to the presence of multiple species.}} 

    \begin{tabular}{lcccccc}
        \toprule

         Cation&$m$ ($\mu_B$) & $T_H$ (K) &   $J_{AB}$ (K) & $J_{BB}$ (K)\\
         \hline
         Cr& 3.0 & \textcolor{black}{87(4)} & \textcolor{black}{-22(3)} & \textcolor{black}{-18(5)} \\
         Mn& 3.0 & \textcolor{black}{181(3)} &  \textcolor{black}{-40(4)} & \textcolor{black}{12(2)} \\
         Fe $T_d$& -5.0 & \textcolor{black}{147(8)} &  \textcolor{black}{-58(4)} & - \\
         Fe $O_h$& 5.0 & \textcolor{black}{100(4)} &  \textcolor{black}{-28(3)} & \textcolor{black}{-27(5)} \\
         Co& -4.0 & \textcolor{black}{172(6)} &   \textcolor{black}{-56(3)}& - \\
         Ni& 2.2 & \textcolor{black}{173(7)} &  \textcolor{black}{-81(4)}  & \textcolor{black}{27(2)} \\
         \toprule
         Probe &$m_{obs}$ ($\mu_B$/f.u.)& $T_H$ (K) & &\\
         \hline
         XMCD & \textcolor{black}{1.7(2)} & \textcolor{black}{89(1)} &   & \\
         SQUID& \textcolor{black}{1.87(5)} & \textcolor{black}{94(1)} &   & \\
         \toprule
    \end{tabular}
    \label{table:critexponents}
\end{table}

\subsection{Nature of the magnetic phase transition in \ch{(Cr,Mn,Fe,Co,Ni)3O4}}

\begin{figure}[tb]
\includegraphics{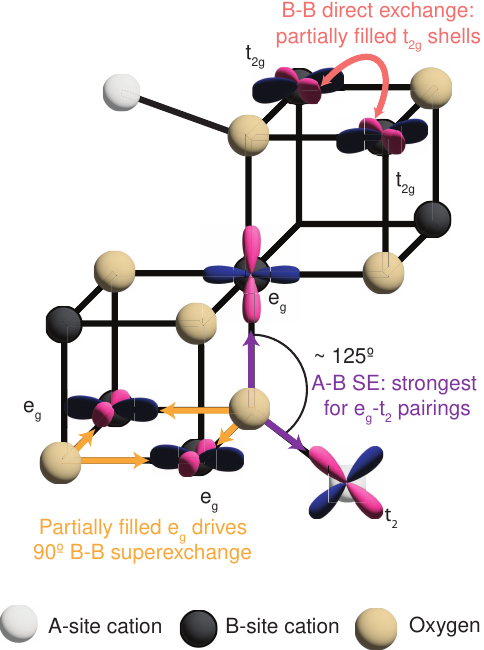}
\caption{\textbf{Main magnetic exchange pathways available in \ch{(Cr,Mn,Fe,Co,Ni)3O4}.} Spinel unit cell fragment, emphasizing the most prominent exchange pathways available in this structure –– $A$-$B$ superexchange, $B$-$B$ superexchange, and $B$-$B$ direct exchange –– and the orbitals which most strongly participate in each kind of interaction. Adapted from Ref.~\cite{Goodenough_spinel_interactions}. } 
 \label{fig:power_law}
\end{figure}

The overall richness in magnetic behavior of \ch{(Cr,Mn,Fe,Co,Ni)_3O4} can be understood by considering the available magnetic interactions in the spinel structure, and their dependence on \textit{3d} subshell filling. Magnetism in \textit{3d} transition metal spinels is dominated, to first order, by three magnetic exchange pathways~\cite{Goodenough_spinel_interactions}, which are depicted in Fig~\ref{fig:power_law}: $A$-$B$ superexchange, which is mediated by corner-shared oxygen between $A$ and $B$ cations at a 120\degree angle, $B$-$B$ superexchange, mediated by the two edge-shared oxygens that connect $B$ cations, and $B$-$B$ direct exchange, which occurs due to direct \textit{3d} overlap between $B$ cations. 

Due to symmetry constraints, each of these interactions is maximized by specific \textit{3d} subshells. $A$-$B$ and $B$-$B$ superexchange rely on overlap between the respective cations and their shared oxygens. For $B$ ($A$) sites, orbital overlap (and hence the magnitude of magnetic exchange) is maximal for $e_g$ ($t_2$) orbitals, which naturally point in the direction of the ligands (and hence the oxygen \textit{p} orbitals), resulting in bonding with $\sigma$ character. For $t_{2g}$ ($e$) orbitals, they instead point in between the ligands, so the bonding acquires $\pi$-like character, and superexchange is weaker, although nonvanishing. Finally, as direct exchange relies on direct \textit{3d} orbital overlap across the relatively small $B$-$B$ cation distance, it is only possible for $t_{2g}$ orbitals that naturally point across the shared edge. 

Ferrimagnetism in spinels –– as in \ch{(Cr,Mn,Fe,Co,Ni)3O4} –– arises from antiferromagnetic $A$-$B$ coupling, whilst $B$-$B$ superexchange tends to be ferromagnetic, and $B$-$B$ direct exchange favors antiferromagnetism to minimize electron repulsion.  All of these interactions are maximized at half filling for each \textit{3d} subshell, and vanish when at full filling. 

These observations allow us to rationalize the elementally-resolved temperature-dependent magnetic moments presented in Fig.~\ref{fig:momenta} (b,e): $A$-site cations (\ch{Fe^{3+}} $T_d$ and \ch{Co^{2+}}) are only subjected to AFM $A$-$B$ superexchange, with no competing interactions, so at the onset of the ordering transition, they harden the fastest. Furthermore, they both present a half-filled $t_{2g}$ shell, maximizing $A$-$B$ superexchange. Meanwhile, the $B$-site cations can be subjected to both ferromagnetic ($B$-$B$ superexchange) and antiferromagnetic ($B$-$B$ direct exchange) interactions. The former synergizes with $A$-$B$ superexchange, while the latter opposes it, resulting in a degree of magnetic frustration. However, \ch{Ni^{2+}}, with its filled $t_{2g}$ shell, does not experience significant $B$-$B$ direct exchange.%

On the other end of the spectrum, we find \ch{Cr^{3+}}. With a $t_{2g}^3$ electron configuration, it can only engage in $A$-$B$ (and $B$-$B$) superexchange with the much weaker $\pi$-like channel, and it maximizes $B$-$B$ direct exchange, a set of conditions that maximizes frustration. Similarly, \ch{Fe^{3+}} $O_h$ also presents a half-filled $t_{2g}$ subshell, although in conjunction with half-filled $e_g$ orbitals, which participate in $A$-$B$ superexchange through the stronger $\sigma$ channel. While \ch{Mn^{4+}} is isoelectronic to \ch{Cr^{3+}}, it also presents a significantly smaller ionic radius than any other cation in this structure, which reduces the influence of $B$-$B$ direct exchange. In the case of \ch{Mn^{3+}} $D_{4h}$, its strong Jahn-Teller activity disrupts direct exchange by distorting the sites around it, while presenting a half-filled $e_g$-like orbital, along with an empty one. 

The connection between electron filling and magnetism explains the hierarchy of the different magnetic moments in \ch{(Cr,Mn,Fe,Co,Ni)3O4}: with no competing interactions, \ch{Fe^{3+}} $T_d$, \ch{Ni^{2+}} $O_h$, and \ch{Co^{2+}} $T_d$ order the fastest. For the rest of the cations, all located in the $B$-site the rate of hardening is dictated by their degree of susceptibility to $B$-$B$ direct exchange. These observations also explain the correlations between  \ch{Fe^{3+}} $T_d$ and \ch{Ni^{2+}} $O_h$ near $T_C$ –– both present the most favorable conditions to order for their respective sublattices, driving order through their strong $A$-$B$ superexchange –– and the low-temperature ones between \ch{Cr^{3+}} and \ch{Fe^{3+}} $O_h$, which strongly engage in disruptive $B$-$B$ superexchange.

\begin{figure*}[htbp]\includegraphics[width=\textwidth]{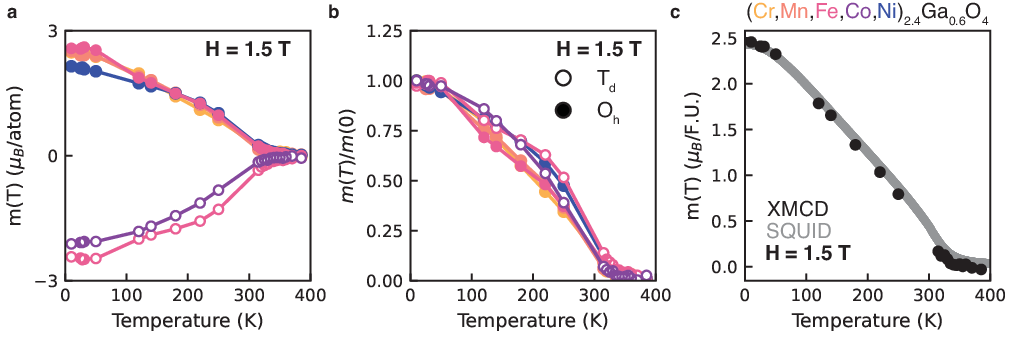}
\caption{\textbf{Cation specific magnetic moments in \ch{(Cr,Mn,Fe,Co,Ni)_{2.4}Ga_{0.6}O4}. a} Temperature-dependent, element-specific XMCD magnetic moments for \ch{(Cr,Mn,Fe,Co,Ni)_{2.4}Ga_{0.6}O4}. In this composition, all of the octahedral sublattice moments have similar magnitudes and rates of evolution. \textbf{b} Element-specific magnetic moments in \ch{(Cr,Mn,Fe,Co,Ni)_{2.4}Ga_{0.6}O4}, as a function of temperature, normalized relative to their base temperature values. All of the cations present near-identical temperature-dependent magnetic moments. \textbf{c} Comparison of the total magnetic moment in \ch{(Cr,Mn,Fe,Co,Ni)_{2.4}Ga_{0.6}O4} as determined \textit{via} XMCD against that observed in SQUID magnetometry under equivalent conditions.}
 \label{fig:momenta20Ga}
\end{figure*}

Further support for these insights is attained by extracting the molecular field around each cation from its elementally-resolved magnetic moment. For a magnetically ordered cation in a mean field picture, the molecular field around it at a given temperature, $H_{ex}(T)$, can be extracted by equating its normalized magnetic moment to the Brillouin function: $m(T)/m_{max} = B_J(x)$, where $x=g\mu_BJH_{tot}/k_BT$, $g$ is the gyromagnetic ratio, $J$ the total angular momentum, and $H_{tot}=H_{applied}+H_{ex}$ is the total magnetic field perceived by the cation. By finding the value of $x$ at each temperature, and using each species' angular momentum state from XMCD, the temperature dependence of the molecular field can be extracted. These results can then be connected to each cation's magnetic interactions:

\begin{equation}
\left\{
\begin{aligned}
    H_{ex,A}(T) &= \lambda_{AB}m_B(T) \\
    H_{ex,B}(T)  &=  \lambda_{AB}m_A(T) + \lambda_{BB}m_B(T)
\end{aligned}
\right.
\end{equation}

\noindent where $\lambda_{AB}$ is a proportionality constant that relates the exchange field felt at each cation site due to the magnetization of the opposite sublattice, and $\lambda_{BB}$ accounts for the interaction with $B$-site neighbors. Within a mean field theory regime, $\lambda_{AB}$ and $\lambda_{BB}$ are expected to be temperature-independent, so that when plotting $H_{ex}$ against the opposite sublattice's magnetization, a linear regime can be identified. 

The extracted values of $H_{ex}$ for each cation are presented in Fig.~\ref{fig:momenta}(f), where they are plotted against the opposite sublattice's magnetization. At lower temperatures (higher sublattice magnetization), an inflection in $H_{ex}$ is observed regardless of chemical species, a behavior that onsets at $\approx120\;K$, and coincides with the saturation of the magnetization in the bulk magnetometry. Therefore, we determine that this behavior is a consequence of magnetic fluctuations becoming significant, and dominating at lower temperatures. This regime was excluded from the calculations for the extraction of the molecular field constants. 

The extracted values for $\lambda_{AB}$ and $\lambda_{BB}$ can then be connected to the average exchange constant felt by each atom due to the $A$ and $B$ sublattices, $J_{AB}, \;J_{BB}$ by considering a Heisenberg Hamiltonian ($\hat{H} = -\sum_{\langle i,j\rangle} J_{ij} \bm{S_i}\bm{Sj}$), and comparing it to the energy felt by each ion due to the exchange field, such that:

\begin{equation}
    J_{ij} = \frac{n_jg_ig_j\mu_B^2}{2z_{ij}} \frac{a^3N_A}{8} \lambda_{ij}
\end{equation}

\noindent where $n_j$ is the number density of magnetic ions in the $j$th sublattice, $g_i,\;g_j$ the respective g-factors, $\mu_B$ is the Bohr magneton, $z_{ij}$ the number of $S_j$ nearest neighbors surrounding spin $S_i$, $a$ the lattice parameter, and $N_A$ is Avogadro's number~\cite{dionne2009magnetic}. The extracted values of $J_{AB}$ and $J_{BB}$ for all cations are presented in Table~\ref{table:critexponents}. 

We can connect this molecular field analysis to the magnetic exchange pathways available in \ch{(Cr,Mn,Fe,Co,Ni)3O4}. From Fig.~\ref{fig:momenta}(f), we can clearly identify that at smaller opposite sublattice magnetizations (corresponding to higher temperatures), Fe $T_d$, Co, and Ni experience significantly larger molecular fields than Cr and Fe $O_h$, with Mn lying between the two extremes, consistent with the rates of ordering of the different cations near $T_C$. Moving onto the extracted exchange constants, we note that for Cr and Fe $O_h$, both $J_{AB}$ and $J_{BB}$ favor antiferromagnetic alignments and are of comparable magnitudes, reinforcing the idea that the interaction between Cr and Fe at the octahedral sites is strongly antiferromagnetic due to direct exchange. Meanwhile, Fe and Co present comparable values of $J_{AB}$, while Mn and Ni present a negative $J_{AB}$ together with a positive $J_{BB}$, indicating that frustration does not play a significant role for these cations. Taken together, the temperature dependence of $H_{ex}$ from Fig.~\ref{fig:momenta}(\textbf{f}), and the extracted values of $J$ support the physical picture of magnetism in \ch{(Cr,Mn,Fe,Co,Ni)3O4} constructed in terms of \textit{3d} orbital symmetry and occupation across the different crystallochemical species present in the material.

\subsection{Effect of non-magnetic substitution on magnetic order in \ch{(Cr,Mn,Fe,Co,Ni)_{2.4}Ga_{0.6}O4}}

Extending our study to the Ga-substituted \ch{(Cr,Mn,Fe,Co,Ni)_{2.4}Ga_{0.6}O4} spinel, we can explore the effect of non-magnetic substitution on high entropy magnetism. \textcolor{black}{It should be emphasized that Ga is not a simple dilutant, as it not only introduces a non-magnetic species but it also modifies the oxidation state of the remaining cations due to its stable oxidation state of \ch{Ga^{3+}}, which is higher than the spinel structure's average of 2.67.} Like its parent compound, \ch{(Cr,Mn,Fe,Co,Ni)_{2.4}Ga_{0.6}O4} is a slightly canted ferrimagnet~\cite{johnstone2022entropy}, with a transition temperature in the vicinity of 300 K. Therefore, the methodology applied to \ch{(Cr,Mn,Fe,Co,Ni)3O4} can be straightforwardly extended here. The full temperature-dependent XAS and XMCD as well as representative LFMT fits are presented in Fig. S4 and Fig. S5.  
Briefly, upon the introduction of Ga, \ch{Cr^{3+}} and \ch{Ni^{2+}} remain fully $O_h$, whilst \ch{Fe^{3+}} exhibits a larger $O_h$ fraction, and \ch{Co^{2+}} goes from a fully $T_d$ coordination, to exhibiting partial $O_h$ occupation. Finally, \ch{Mn} changes both site occupancies and oxidation state, from a nearly entirely $O_h$ to a predominantly $T_d$ site distribution, and a notable reduction from a 3.4+ to a 2.4+ oxidation state.

The resulting XMCD magnetometry for the magnetic cations in \ch{(Cr,Mn,Fe,Co,Ni)_{2.4}Ga_{0.6}O4} is shown in  Fig.~\ref{fig:momenta20Ga}(a) with the lowest temperature (25 K) magnetic moments  ($m_{obs}$) presented in Table~\ref{table:critexponents20Ga}. The low temperature magnetic moments of Mn and Ni remain close to their expected values, 
while moments of 2.5 $\mu_B$ and -2.1 $\mu_B$ for Cr and Co, respectively, are smaller than those corresponding to their fully polarized values. In the case of Fe, the magnetic moments for \ch{Fe^{3+}} in $T_d$ and $O_h$ coordination are nearly identical throughout the measurement range, which coupled with the $O_h$:$T_d$ ratio of 56:44 from XAS/XMCD speciation, results in a small net magnetic moment for Fe, highlighting the importance of extracting the site-specific temperature dependence of their magnetic moments. 

To better compare the magnetic response of cations, we show the cation-resolved magnetic moments normalized to their base temperature values in Fig.~\ref{fig:momenta20Ga}(b). Strikingly, the magnetic moments for the six cations are essentially equivalent both around and away from $T_{\textrm{C}}$, resembling a true ``collective magnetism'' picture, unlike what was observed for the non-diluted parent compound. Unlike in the fully magnetic composition, \ch{Cr^{3+}} and \ch{Fe^{3+}} $O_h$ grow at a rate comparable to the other species in \ch{(Cr,Mn,Fe,Co,Ni)_{2.4}Ga_{0.6}O4}. The agreement between the total XMCD magnetic moment and SQUID magnetometry, shown in Fig.~\ref{fig:momenta20Ga}(c) is also excellent –– all relevant qualitative features are captured, including a broad magnetic transition and the saturation of the magnetic moment below 100 K. This qualitative assessment is validated by studying $T_H$ under the same definition as above (85\% of the low temperature magnetic moment for each cation), which are reported in Table~\ref{table:critexponents20Ga}. Overall, for \ch{(Cr,Mn,Fe,Co,Ni)_{2.4}Ga_{0.6}O4}, there is a much narrower spread in $T_H$. 

\begin{table}
    \centering
\caption{Cation-resolved XMCD magnetic moments ($m_{obs}$) at 25 K, their calculated spin-only values ($m_{calc}$) deduced from the fitted XAS speciation and hardening temperatures ($T_H$) in \ch{(Cr,Mn,Fe,Co,Ni)_{2.4}Ga_{0.6}O4}.}
\label{tab:placeholder}
    \begin{tabular}{ccccc}
        \toprule

         Cation& $m_{obs}$  ($\mu_B$) & $T_H$ (K)\\
         \hline
         Cr& 2.5 & \textcolor{black}{120(4)}\\
         Mn& 2.5 & \textcolor{black}{103(3)}\\
         Fe\textsubscript{$T_d$}& 5.0 & \textcolor{black}{110(3)}\\
         Fe\textsubscript{$O_h$}& 5.0 & \textcolor{black}{86(2)}\\
         Co& -2.1 & \textcolor{black}{139(3)} \\
         Ni& 2.1& \textcolor{black}{123(5)}\\
                  \toprule
         Probe& $m_{obs}$  ($\mu_B$/f.u.) & $T_H$ (K)\\
         \hline
         XMCD & \textcolor{black}{2.5(2)} & \textcolor{black}{89(2)}\\
         SQUID & \textcolor{black}{2.43(7)} & \textcolor{black}{77(1)}\\
         \toprule
    \end{tabular}
    
    \label{table:critexponents20Ga}
\end{table}

Overall, our findings for \ch{(Cr,Mn,Fe,Co,Ni)_{2.4}Ga_{0.6}O4} highlight how a modest compositional change in HEOs can dramatically impact their magnetic phenomenology. Near $T_{\textrm{C}}$, the element-specific magnetic moments exhibit comparable magnitudes and scaling behavior, with no cation dramatically dominating over the others. Furthermore, magnetic hardening in \ch{(Cr,Mn,Fe,Co,Ni)_{2.4}Ga_{0.6}O4} is effectively chemically homogeneous, unlike the fully magnetic \ch{(Cr,Mn,Fe,Co,Ni)_{3}O4}. These differences can be directly connected to the site distribution of Ga in the material. For this discussion, one must remember that \ch{Ga^{3+}} has a fully occupied \textit{d} shell, so it weakens any interactions it engages in. In \ch{(Cr,Mn,Fe,Co,Ni)_{2.4}Ga_{0.6}O4}, Ga occupies 24\% of all available $O_h$ sites, and 12\% of $T_d$ sites. As Ga is mainly located in the $O_h$ sublattice, it can disrupts specific magnetic interactions in the spinel structure: for cations in $T_d$ sites, it weakens the main \textit{A}-\textit{B} superexchange interaction. This results in the lower moment observed for \ch{Co^{2+}} $T_d$, while \ch{Fe^{3+}} remains largely unaffected, as it has more available orbitals to interact with the B sublattice through its half-filled  $t_2$ and $e$ shells, mitigating the effect of broken magnetic linkages. For cations occupying $O_h$ sites, Ga weakens both types of \textit{B}-\textit{B} exchange interactions -- direct exchange and superexchange. However, Ga dilution affects \textit{B}-\textit{B} direct exchange more dramatically than \textit{B}-\textit{B} superexchange. This happens because the latter interaction's effect is supplemented by \textit{A}-\textit{B} superexchange, which is barely altered due to the smaller Ga $T_d$ concentration (11\%). Meanwhile, the broken magnetic linkages automatically remove 24\% of the \textit{B}-\textit{B} direct exchange channels. This combination of circumstances results in a reduction in the degree of frustration for Cr and Fe $O_h$ in \ch{(Cr,Mn,Fe,Co,Ni)_{2.4}Ga_{0.6}O4}.

\section*{Discussion}

We have successfully used XMCD to disentangle cation-resolved magnetic moments in two spinel high entropy oxide systems –– \ch{(Cr,Mn,Fe,Co,Ni)3O4} and \ch{(Cr,Mn,Fe,Co,Ni)_{2.4}Ga_{0.6}O4} –– allowing us to understand how emergent behavior arises in HEOs. This procedure returns reasonable magnetic moment values for all of the constituents, and when summed, these moments reconstruct the bulk magnetic response.
The cation-specific magnetic moments reveal that significant changes in magnetic phenomenology arise with relatively modest compositional tweaks: while in both cases the magnetic transition occurs simultaneously for all cations, their temperature evolutions differ significantly. In \ch{(Cr,Mn,Fe,Co,Ni)3O4}, \ch{Cr^{3+}} and \ch{Fe^{3+}} cations occupying the octahedrally coordinated $B$-sites lag behind the rest of the constituents throughout the majority of the ordered regime. As highlighted in the manuscript, these differences in magnetic behavior can be understood in terms of the filling of the \textit{3d} crystal field energy levels for the different cations, as they dictate their ability to engage in the different magnetic exchange channels available in the spinel structure. For the aforementioned cations, the interplay between AFM $A$-$B$ superexchange, FM $B$-$B$ superexchange, and AFM $B$-$B$ direct exchange results in magnetic frustration. Meanwhile, for \ch{(Cr,Mn,Fe,Co,Ni)_{2.4}Ga_{0.6}O4}, the introduction of \ch{Ga^{3+}} as a non-magnetic substitute significantly relieves magnetic frustration on these ions by reducing the number of constraints their moments must fulfill.

Overall, our findings show that, to design magnetism in HEO spinels, one needs intricate knowledge of not just their site selectivities, but the details of their available exchange pathways, and the magnetically active orbitals involved in each of them. \textcolor{black}{These observations are valuable not just from a fundamental point of view, but they are also relevant for the eventual implementation of spinel HEOs in applied settings. 
Together with previously derived guidelines for the control of site selectivity and magnetic properties in HEOs~\cite{johnstone2022entropy}, the selection of cations that tend towards specific orbital fillings allows one to tailor their magnetic response to fulfill specific criteria. For example, in magnetocaloric cooling, one targets broad magnetic transitions near the desired operating window. In HEOs, these kind of windows could be engineered by tuning both the cations' magnetic interaction strengths -- modifying the Curie temperature -- and the degree of overall magnetic frustration~\cite{Zhitomirsky_2003_frustrated_magnetocaloric}, by increasing the fraction of octahedral cations that engage in the unfavourable $e_g-$O$-e_g$ exchange channel. Together, these effects should result in broader windows of magnetocaloric operation. Alternatively, the knowledge acquired here can also be translated to electrocatalysis settings, an area where, in conventional materials, octahedral cations' $e_g$ occupancy has been established as a valuable descriptor of catalytic activity~\cite{Suntivich2011_catalysis}. Together with spinel HEOs' structural stability, additional optimization opportunities arise on this front.}

\section*{Methods}
\subsection*{Synthesis of spinel HEO powders}

Spinel HEO powders with nominal stoichiometries \ch{(Cr,Mn,Fe,Co,Ni)_{3-x}Ga_xO4} (x = 0, 0.6) were synthesized by adapting a previously reported solution combustion synthesis route~\cite{JACS-synthesis-dep}. Briefly, stoichiometric amounts of the corresponding metal nitrates were dissolved in water, then stirred for 10 minutes. At this point, glycine was added to the mixture to achieve a fuel-to-oxidizers ratio ($\phi$) of 1, as described by the following reaction:

\begin{math}
    Me^{\nu +}(\text{NO}_3)_{\nu} + \frac{5}{9}\phi \nu \text{C}_2\text{H}_5\text{NO}_2 +\frac{5}{4} \nu (\phi-1) O_2 \rightarrow Me^{\nu +}\text{O}_{\nu/2} + \left(\frac{5\phi+9}{18} \right) \nu \text{N}_2 + \frac{25}{18}\nu \phi \text{H}_2 + \frac{10}{9} \nu \phi \text{CO}_2
\end{math}

\noindent where $Me^{\nu +}(\text{NO}_3)_{\nu}$ refers to a $\nu$-valent metal cation. The solution was then heated under constant stirring to 80 \degree C to trigger nitrate-glycine complexation, and kept at that temperature until a viscous gel was formed. Combustion was then achieved by placing the gels on a hot plate preheated to 380 \degree C. The resulting powders were ground and annealed at 1000 \degree C for 24 hours, and quenched in air.

\subsection{Powder x-ray diffraction}

Powder diffraction measurements were conducted using a Bruker D8 Advance diffractometer equipped with a copper source and Johansson monochromator, resulting in a monochromatic beam wavelength of $\lambda_{K_{\alpha_1}}$ = 1.5046 \AA. The instrument is equipped with a LYNXEYE XE-T silicon strip detector that enables it to efficiently filter background fluorescence, which is significant for the elements under study. Measurements were carried out over a range of 1 < $Q$ < 6 \AA$^{-1}$, with a step size of 99.7 n\AA$^{-1}$. Rietveld refinements for all samples were performed using the FullProf Suite \cite{RodriguezCarvajal-Fullprof}, and are presented in Fig. S1. All measurements were performed on finely ground powders –– sifted through a 400 mesh –– to minimize artifacts.

For \ch{(Cr,Mn,Fe,Co,Ni)3O4} only, medium-resolution synchrotron powder diffraction measurements were conducted out at beamline 28-ID-1 (PDF) of the National Synchrotron Light Source II (Brookhaven National Laboratory, Upton, NY). The powders were placed inside polyimide capillary tubes. An incident wavelength of 0.1665 \AA \ was used, and diffraction data were collected using an amorphous silicon area detector (PerkinElmer, Waltham, MA), placed 216.7 mm behind the samples. Data collection proceeded at room temperature. Calibration of the experimental setup was done using a nickel powder beamline standard.

\subsection{X-ray absorption spectroscopy}

Temperature-dependent x-ray absorption spectroscopy (XAS) and x-ray magnetic circular dichroism (XMCD) experiments were conducted at the absorption endstation of beamline I10-1 at Diamond Light Source, and at the elastic endstation of the REIXS beamline of the Canadian Light Source~\cite{REIXS-beamline}.

Experiments were carried out at normal incidence, under an applied magnetic field normal to the sample surface. At I10, an applied field of 1.50 T was achieved \textit{via} an electromagnet. Meanwhile, at REIXS, an applied field of approximately 0.6 T (at the sample's surface) was obtained by mounting the samples on top of two stacked \ch{NdFeB} permanent magnets. 

Spectra were collected by measuring the drain current while scanning the incident photon energy across the transition metal $L_{2,3}$ edges, thus obtaining total electron yield (TEY) data. The samples were mounted on silver paint to improve thermal contact. 

An additional set of room temperature XAS data was collected at the inelastic endstation of the REIXS beamline at the Canadian Light Source. This data was collected in inverse partial fluorescence yield (IPFY) mode, using the spectrometer to isolate the O K emission line for Cr and Mn, and a Si drift detector for Fe, Co, and Ni.

\subsection{Magnetometry measurements}

Magnetic moment data was collected using a Quantum Design MPMS-3 SQUID magnetometer. The samples were mounted in polypropylene capsules, and loaded into a brass sample holder. Temperature-dependent M/H curves were collected between 2 and 400 K, under applied fields H equivalent to those available at ID-10 (1.5 T). Data was collected under both zero field cooled (ZFC) and field cooled (FC) protocols. However, as the applied fields are larger than the material coercive field, the ZFC and FC curves were effectively identical. Low-field magnetic susceptibility data were also collected to approximate the zero applied field ordering behavior, and are presented in Fig. S2.

\section*{Data availability}
The authors declare that the main data supporting the findings of this study are available within the paper and its Supplementary Information files. Raw XAS data used in this study have been deposited in the Zenodo database under the digital object identifier: doi.org/10.5281/zenodo.19493154.
Additional data are available from the authors upon request.\\

\section*{Acknowledgments}
The authors thank Austin Ferrenti for helpful discussions surrounding the XMCD data analysis, and Abraham Mancilla, Lucas Korol, and Grant Harris for their assistance with data collection. The authors thank A. Tanaka for making his XTLS 9.0 code available. The authors thank Dr. Milinda Abeykoon, Dr. Solveig S. Aamlid, and Allison Pavlik for help carrying out the synchrotron diffraction experiments. 
This work was supported by the NSERC-NSF Designing Materials to Revolutionize and Engineer our Future (DMREF) Alliance Grant in partnership with DMREF Grant Number DMR2523217. This research was undertaken thanks in part to funding from the Max Planck-UBC-UTokyo Centre for Quantum Materials and the Canada First Research Excellence Fund, Quantum Materials and Future Technologies Program. This work was also supported by the Natural Sciences and Engineering Research Council of Canada (NSERC), the Canadian Institute for Advanced Research (CIFAR), and the Sloan Research Fellowships Program. The authors would like to thank Diamond Light Source for beamtime (proposal MM36686), and the staff of beamline I10 for assistance with data collection.
Part of the research described in this paper was performed at the Canadian Light Source, a national research facility of the University of Saskatchewan, which is supported by the Canada Foundation for Innovation (CFI), the Natural Sciences and Engineering Research Council (NSERC), the National Research Council (NRC), the Canadian Institutes of Health Research (CIHR), the Government of Saskatchewan, and the University of Saskatchewan. This research used beamline 28-ID-1 of the National Synchrotron Light Source II, a U.S. Department of Energy (DOE) Office of Science User Facility operated for the DOE Office of Science by Brookhaven National Laboratory under Contract No. DE-SC0012704. \\

\section*{Author contributions}

M.U.G.R. and A.M.H. conceived the project and designed the experimental approach. Materials preparation, characterization, and data analysis were performed by M.U.G.R. with support from M.B., J.F., R.S., T.D.B., P.B., and R.J.G. in the XAS experiments and guidance from R.J.G., G.A.S., and L.H.T. in the data analysis. C.F.C. and L.H.T. performed the LFMT calculations. The manuscript was written by M.U.G.R. and A.M.H. with input from all authors.

\bibliography{bibliography}

\end{document}